\documentclass[prl,superscriptaddress,aps,twocolumn,showpacs,floatfix,amsmath,amssymb]{revtex4}

\usepackage{graphicx}
\usepackage{dcolumn}
\usepackage{epstopdf}

\newcommand{\ket}[1]{| #1 \rangle}
\newcommand{\bra}[1]{\langle #1 |}
\newcommand{\Id}{\mathbb I}

\begin{document}
\title{Optimal correlations in many-body quantum systems}

\author{L. Amico}
\affiliation{CNR-MATIS-IMM $\&$ Dipartimento di Fisica e Astronomia Universit\'a di Catania, C/O ed. 10, viale A. Doria 6,
  95125 Catania, Italy}

\author{D. Rossini}
\affiliation{NEST, Scuola Normale Superiore $\&$ Istituto Nanoscienze-CNR, Piazza dei Cavalieri 7, I-56126 Pisa, Italy}

\author{A. Hamma}
\affiliation{Perimeter Institute for Theoretical Physics, 31 Caroline St. N, Waterloo ON, N2L 2Y5, Canada}

\author{V. E. Korepin}
\affiliation{C. N. Yang Institute for Theoretical Physics, State University of New York at Stony Brook, NY 11794-3840, USA}

\begin{abstract}
  Information and correlations in a quantum system are closely related through the process of measurement. 
  We explore such relation in a many-body quantum setting, effectively bridging 
  between quantum metrology and condensed matter physics.
  To this aim we adopt the information-theory view of correlations, and study the amount of correlations 
  after certain classes of Positive-Operator-Valued Measurements are locally performed. 
  As many-body system we consider a one-dimensional array of interacting two-level systems (a spin chain)
  at zero temperature, where quantum effects are most pronounced. 
  We demonstrate how the optimal strategy to extract the correlations depends on the quantum phase 
  through a subtle interplay between local interactions and coherence.
\end{abstract}



\maketitle

The relation between correlations and measurement is well known in quantum metrology, 
where the optimal measurement strategy 
to extract information has been thoroughly investigated~\cite{info-complete,qmetrology}.
In that context Positive-Operator-Valued Measurements (POVMs) and Informationally-Complete (IC) 
measurements are of particular importance since, contrary to simple projective measurements, 
they allow a complete tomography of the quantum state~\cite{peres_povm}. 
They have been playing an important role also
in foundational aspects of quantum mechanics, quantum information science, as well as 
in the physics of dissipative systems~\cite{sic, sic-povm, foundations, tomography}. 

In this Letter we aim at establishing a bridge between quantum metrology and quantum many particles physics.
We consider subsystems $A$ and $B$ in a many-body ground state, and analyze the correlations 
resulting from POVMs performed on one of them, say $B$. 
Emphasis will be devoted to the {\it optimal correlations}, namely the maximal amount 
of correlations established between $A$ and $B$.
We observe that performing a POVM on a given physical system is equivalent 
to performing a projective measurement on an enlarged system where the original one is coupled 
with a given ``ancilla'' (Naimark's theorem). Such an ancilla can be an appropriate subsystem, 
and then analyzing correlations induced by a POVM on a local degree of freedom, say $B$, 
is an effective way to study correlations of higher order (spin-block correlations). 
Equivalently, the ancilla can be a suitable environment entangled to the system, 
and then correlations can give precious information on the decoherence 
of the state of the local constituent $A$. 

The total amount of correlations in any bipartite (mixed) 
quantum state $\hat{\rho}_{AB}$ is given by the {\it mutual information}:
${\cal I}_{AB} \equiv {\cal S}(\hat{\rho}_A) + {\cal S}(\hat{\rho}_B) - {\cal S}(\hat{\rho}_{AB})$,
where ${\cal S}(\hat{\rho}) = - {\rm Tr} [\hat{\rho} \, {\rm log}_2 \hat{\rho}]$ is the von Neumann entropy.
A central quantity we will address below is the quantum  {\it conditional entropy} ${\cal S_C}$, 
quantifying the ignorance about the composite system $AB$, once subsystem $B$
has been measured with a generic POVM $\{ \hat{B}_k \}$:
\begin{equation}
  {\cal S_C}(\hat{\rho}_{AB} | \{\hat{B}_k\}) = \sum_k p_k S(\hat{\rho}_{AB}^{(k)}) \,.
  \label{eq:QCond}
\end{equation}
Here $\hat{\rho}_{AB}^{(k)} $ denotes the state of the composite system $AB$, conditioned to a given outcome 
of $\hat{B}_k$:  $\hat{\rho}_{AB}^{(k)} = \frac{1}{p_k} [(\hat{\Id} \otimes \hat{B}_k) \, \hat{\rho}_{AB}]$ 
with $\hat{\Id}$ denoting the identity operator on the subsystem $A$ and 
$p_k = {\rm Tr} [(\hat{\Id} \otimes \hat{B}_k) \, \hat{\rho}_{AB}]$. 
The global amount ${\cal C}_{AB}$ of optimal (classical) correlations between constituents $A$ and $B$ 
is established after applying a set of measurements on ${B}$ that disturbs the least the part $A$:
\begin{equation}
  {\cal C}_{AB} = \max_{\{\hat{B}_k\}} \big[ {\cal S}(\hat{\rho}_A) - {\cal S_C} (\hat{\rho}_{AB} | \{\hat{B}_k\}) \big] \,,
  \label{eq:Class}
\end{equation}
where the maximization is with respect to the measurement strategy. 
The quantum discord~\cite{qdiscord} quantifies the amount of quantum correlations
and is defined as the difference between total and classical ones:
${\cal Q}_{AB} = {\cal I}_{AB} - {\cal C}_{AB}$. 
The maximization in Eq.~\eqref{eq:Class} is generally a daunting task, since the optimization procedure 
has to be performed on the whole set of possible POVMs. 

We apply the above notions to the case where $A$ and $B$ are individual spins 
of a quantum spin chain, and consider both von Neumann projective measurements $\hat{M}_{proj}$, 
and generalized POVMs $\hat{M}_{povm}$~\cite{note}. 
We design a strategy to exploit the information input given by the 
physical system hosting the two spins. Namely, we assume that the symmetry of the POVM 
is fixed by the symmetry of the {\it local interactions} occurring in the physical system. 
However, we shall see this is not enough to optimize correlations,
as the {\it coherence} of the many-body system is going to play a major role.

{\it Models and measurements.}---
As many body system we consider a one dimensional array of spins $1/2$ interacting anisotropically 
along the three spatial directions with interaction strengths $J_x, \, J_y, \, J_z$, 
and subjected to a uniform external field $h$.
The Hamiltonian reads 
\begin{equation}
  \label{general-spin}
        {\cal \hat{H}} = \sum_i\left( J_x \hat{\sigma}_i^x \hat{\sigma}_{i+1}^x  + J_y \hat{\sigma}_i^y \hat{\sigma}_{i+1}^y 
        + J_z \hat{\sigma}_i^z \hat{\sigma}_{i+1}^z \right) - h \sum_i \hat{\sigma}_i^z ,
\end{equation}
where $\hat{\sigma}_i^\alpha$ ($\alpha = x,y,z$) are the Pauli matrices on site $i$.
Hereafter we set $\vert J_x \vert = 1$ as the energy scale and work in units of $\hbar = 1$.
At zero temperature different quantum phases can exist, separated 
by Quantum Phase Transitions (QPTs)~\cite{sachdev}. 
Moreover a completely factorized ground state may occur at a specific value 
of the field $h_f$~\cite{factorization}.
For $xyz$ spin chains of Eq.~\eqref{general-spin} this is
given by: $h_f = 2 \sqrt{(1-J_z) (J_y-J_z)}$~\cite{giampaolo}.

We will discuss Eq.~\eqref{general-spin} in the following cases: 
I) The ferromagnetic Ising chain ($J_y=J_z=0$, $J_x=-1$), which undergoes a QPT
at $h_c=1$, and factorization at $h_f=0$.
It can be experimentally realized with the magnetic compound ${\rm Co \, Nb_2 \, O_6}$~\cite{coldea}. 
II) The non-integrable antiferromagnetic $xyx$ model ($J_x=J_z=1$), with $J_y = 1/4$
(this is the case experimentally realized with ${\rm Cs_2 \, Co \, Cl_4}$~\cite{kenzelmann}).
Such model displays a QPT at $h_c \simeq 3.21$ and factorization at $h_f \simeq 3.16$. 
III) The antiferromagnetic anisotropic $xxz$ Heisenberg chain ($J_x = J_y = 1$) with $J_z = 1/2$.
At zero field it presents a critical $xy$ phase with quasi-long range order (quasi-lro)
for $\vert J_z \vert < 1$; this is separated by two classical phases with QPTs 
at $J_z = \pm 1$. For $h\neq 0$ the $xy$ phase is a strip in the phase diagram, 
eventually turning into polarized phases for sufficiently strong magnetic field 
(the factorization phenomenon degenerates in the saturation occurring as a first order transition).
%
Despite local interactions are clearly different, both the quantum Ising 
and $xyx$ models display an Ising-like QPT with $Z_2$-symmetry breaking; 
the $xxz$ model, instead, is characterized by a critical phase without order parameter.

\begin{figure}[!t]
  \centering
  \includegraphics[width=0.5\columnwidth]{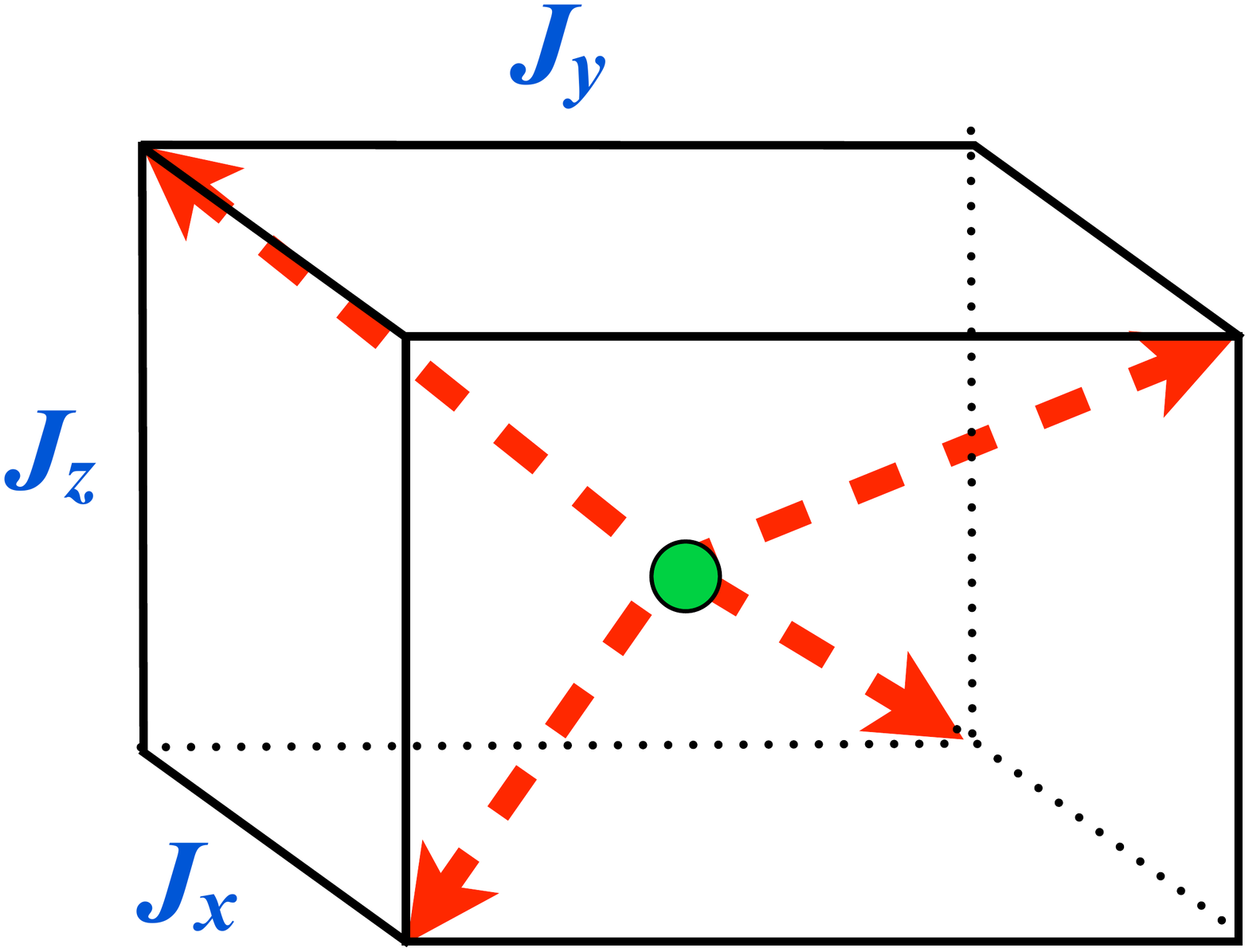}
  \caption{(color online). The four vectors entering the POVM measurement of Eq.~\eqref{eq:POVM}. 
    They point from the center to non-adjacent corners of a parallelepiped 
    with edges fixed by the anisotropic interaction occurring into the system.}
  \label{fig:cube}
\end{figure}

We first deal with standard projective measurements $\hat{M}_{proj} = \{ \hat{B}_\pm \}$ 
along the field direction, defined by $\hat{B}_\pm = \frac{1}{2} (\hat{\Id} \pm \hat{\sigma}^z)$.
Then we engineer a more sophisticated set of POVMs, such that the symmetry of the measurement
keeps track of local interactions between the spins.
Specifically, we look at the interactions $J_x, \, J_y, \, J_z$ 
entering Eq.~\eqref{general-spin}, and design the following $\hat{M}_{povm} = \{\hat{B}_k\}$: 
\begin{equation}
  \hat{B}_k = \frac{1}{4} \left( \hat{\Id} + \vec{a}_k \cdot \hat{\vec{\sigma}} \right) , 
  \qquad k=1 \ldots 4 \,,
  \label{eq:POVM}
\end{equation} 
where $\hat{\vec{\sigma}} = (\hat{\sigma}^x, \, \hat{\sigma}^y, \, \hat{\sigma}^z)$ 
and $\vec{a}_k$ is such that $\vec{a}_1= \alpha (J_x,J_y,J_z)$, $\vec{a}_2=\alpha (J_x,-J_y,-J_z)$, 
$\vec{a}_3=\alpha (-J_x,J_y,-J_z)$, $\vec{a}_4=\alpha (-J_x,-J_y,J_z)$ and
$\alpha^{-1} = \sqrt{J_x^2+J_y^2+J_z^2}$ (see Fig.~\ref{fig:cube}).
%
%
For generic $J_\alpha$, $\hat{M}_{povm} $ will be denoted as Coupling-oriented Informationally Complete (CIC) POVM. 
The choice of the vectors ${\vec a}_k$ in Eq.~\eqref{eq:POVM} reflects 
the symmetry of the Hamiltonian by changing  $J_\alpha \rightarrow -J_\alpha$ 
by $\hat{U} \, {\cal \hat{H}} \, \hat{U}^{\dagger}$, 
with $\hat{U}^\alpha = \prod_i \hat{\sigma}_{2i+1}^\alpha$.
In the isotropic case $J_x = J_y = J_z$, the POVM in Eq.~\eqref{eq:POVM} degenerates 
in a Symmetric Informationally Complete (SIC) POVM~\cite{sic-povm}.
We comment that all the $\hat{B}_k$ do not depend on the external field explicitly. 
We shall see that such a dependence enters in a subtle way related to the macroscopic order of the system.

In order to compute the amount of correlations between any two spins at distance $r \equiv \vert A-B \vert$,
one needs to access the single- and two-spin reduced density matrices $\hat{\rho}_A$ and $\hat{\rho}_{AB}$.
Hereafter we focus on the symmetry-broken ground states of the Hamiltonian in Eq.~\eqref{general-spin}, 
which is symmetric under a global phase flip along the $z$-axis~\cite{palacios}.
We observe that the generic two-site reduced density matrices of such states is beyond 
the so called ``X-state'' structure (emerging in symmetric states, and for which expressions 
for quantum and classical correlations are known~\cite{rau2010}).
Therefore, in the present case, in principle the optimal correlations might be achieved 
beyond projective measurements~\cite{zaraket}.
To access all the required two-point correlators, we resort 
to the density matrix renormalization group approach with open boundary conditions~\cite{dmrg}.
We consider sufficiently long chains, such to reduce unwanted edge effects 
and to approach the ideal thermodynamic limit.
For the Ising and the $xyx$ model, we add a small longitudinal field $h_x \sim 10^{-6}$ 
along the $xy$ plane, to ensure the $Z_2$-symmetry breaking.

\begin{figure}[!t]
  \centering
  \includegraphics[width=0.9\columnwidth]{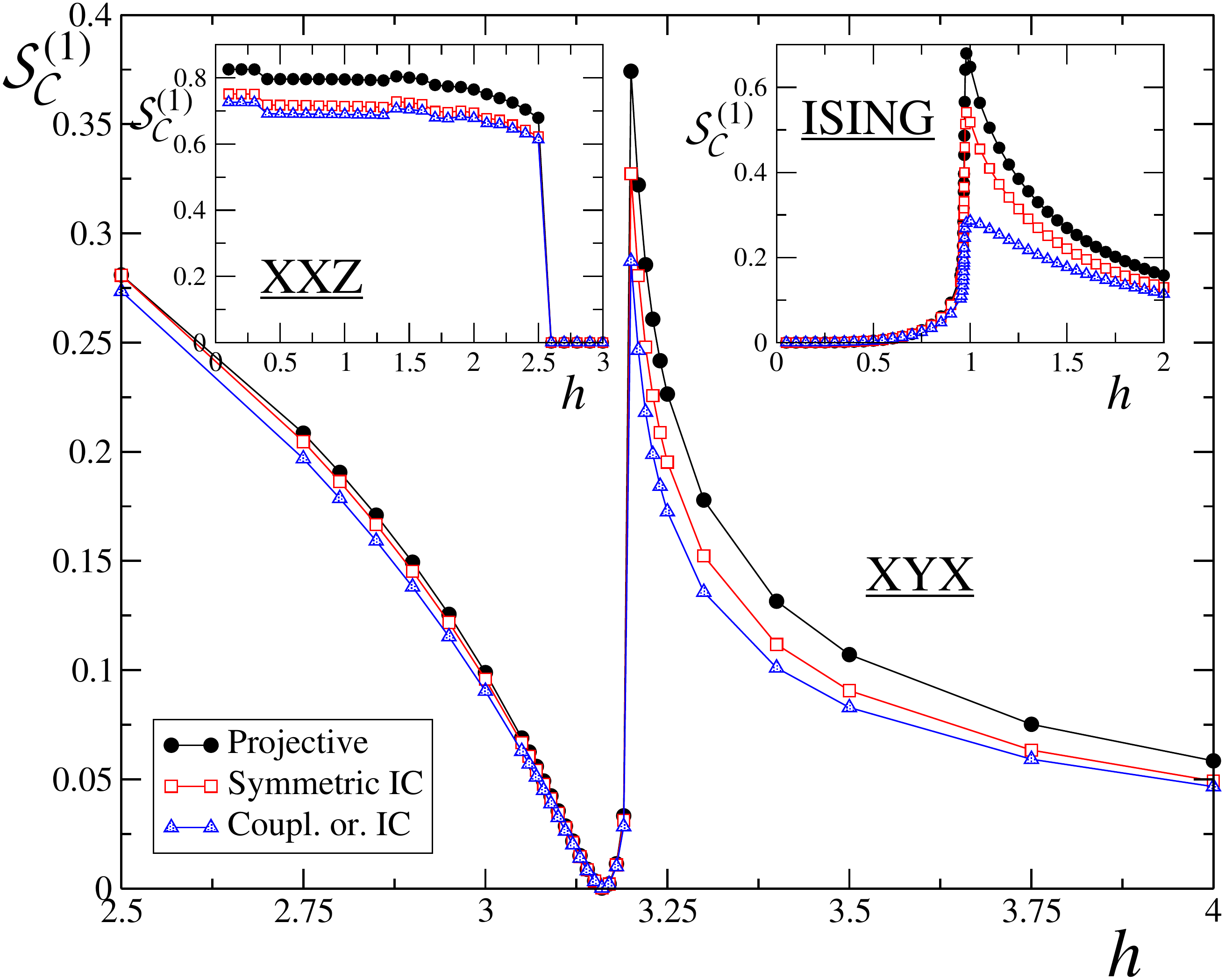}
  \caption{(color online). Conditional entropy ${\cal S}^{(1)}_{\cal C}$ for the three cases: 
    I) Ising, II) $xyx$ and III) $xxz$ model in an external field [see Eq.~\eqref{general-spin}].
    Correlations are considered between two nearest neighbors 
    at the center of the chain ($r=1$); the length of the chain is $L = 200$.
    The various curves correspond to different measurement strategies on subsystem $B$:
    standard projective measurement along the $z$ direction (black circles), SIC POVM (red squares), 
    as well as IC POVM set by the specific interactions of the model (blue diamonds).}
  \label{fig:Cond_Entro}
\end{figure}

{\it Comparison between different measurement strategies.}---
We start our analysis by presenting results obtained for the quantum conditional entropy 
${\cal S}^{(r)}_{\cal C}$ in Eq.~\eqref{eq:QCond}, probing how the local interactions 
affect the measurement, without any further optimization.
Figure~\ref{fig:Cond_Entro} displays ${\cal S}^{(1)}_{\cal C}$ for two neighboring spins
respectively for the Ising model, the $xyx$ model and the $xxz$ model in a transverse field $h$.
The total amount of correlations reflects the main properties of the ground state:
in particular the peaks denote QPT points that are associated to a divergence in their first derivative,
while factorization fields are marked by the vanishing of all correlations.
In all the three considered spin systems, measurement performances decrease 
from the CIC POVM to the SIC POVM and to the projective measurement 
along the computational basis $z$.

Much larger amounts of correlations can be achieved by performing suitable
optimization strategies on the measurements considered above.
In the following we apply two different kinds of optimization:
$i)$ We rotate the direction of the elements $\hat{B}_k$ of the projective measurement $\hat{M}_{proj}$ and 
of the POVM $\hat{M}_{povm}$ on the Bloch sphere, by keeping the angles between the vectors $\vec{a}_j$ 
constant (this corresponds to a rigid rotation of the experimental apparatus); 
optimal correlations are thus obtained by maximizing over the angles $(\theta, \phi)$ entering the rotation. 
$ii)$ In the case of CIC POVM, we independently vary the three parameters $J_x,J_y,J_z$, 
defining the direction of the vectors $\vec{a}_k$ in $\hat{M}_{povm}$ (see Fig.~\ref{fig:cube})~\cite{note2}.

\begin{figure}[!t]
  \centering
  \includegraphics[width=0.9\columnwidth]{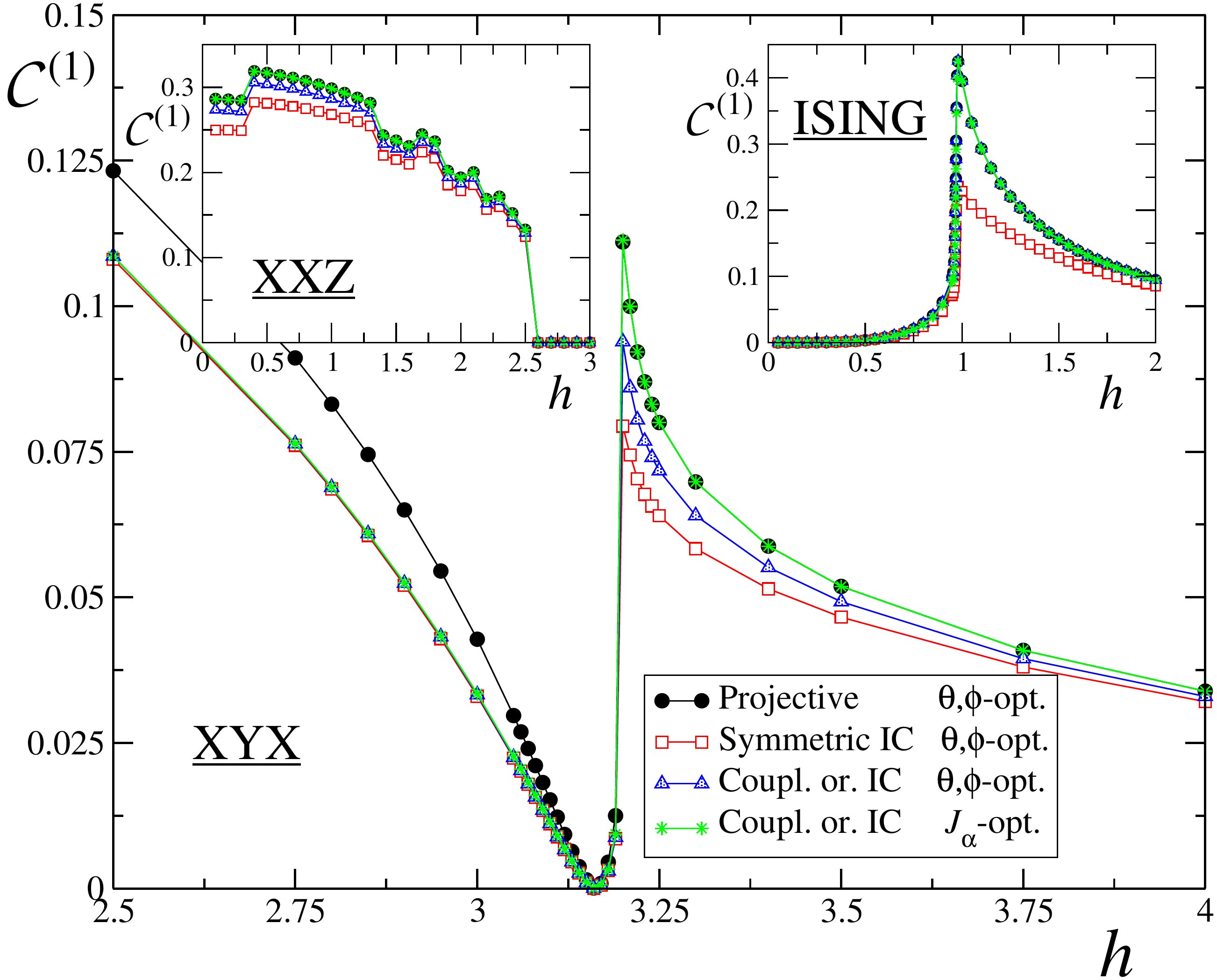}
   \caption{(color online). Same as in Fig.~\ref{fig:Cond_Entro}, but for the classical correlations 
     ${\mathcal C}^{(1)}$ optimized over the given measurement strategy.
     In the first three cases, projective measurement, SIC POVM and IC POVM (see the symbol
     pattern of Fig.~\ref{fig:Cond_Entro}), the optimization in Eq.~\eqref{eq:Class} is performed 
     on the rotation angles $(\theta, \phi)$ of the Bloch sphere of subsystem $B$.
     The green stars refer to IC POVM optimized by varying the three parameters ${J_\alpha}$
     defining the direction of the vectors $\vec{a}_k$.}
   \label{fig:Classical_Corr}
\end{figure}

In Fig.~\ref{fig:Classical_Corr} we display the classical correlations
between two neighboring spins for the three considered models,
by adopting the optimizations discussed above.
Similarly to the quantum conditional entropy, ${\mathcal C}^{(r=1)}$ displays a noticeable 
dependence on the model and on the magnetic field. 
While in the Ising model the CIC angle-dependent POVM and projective measurement give 
the same answer, for the $xyx$ and $xxz$ model 
the angle-dependent strategy is not optimal and is outperformed by the projective measurements. 
The 3-parameters POVM optimizations provide the same correlations of the projective measurements
in the disordered regions and in $xxz$ model; 
however, where the order parameter $\langle \sigma^x\rangle \neq 0$,
they are still worse than the projective measurement.

\begin{figure}[!t]
  \centering
  \includegraphics[width=.8\columnwidth]{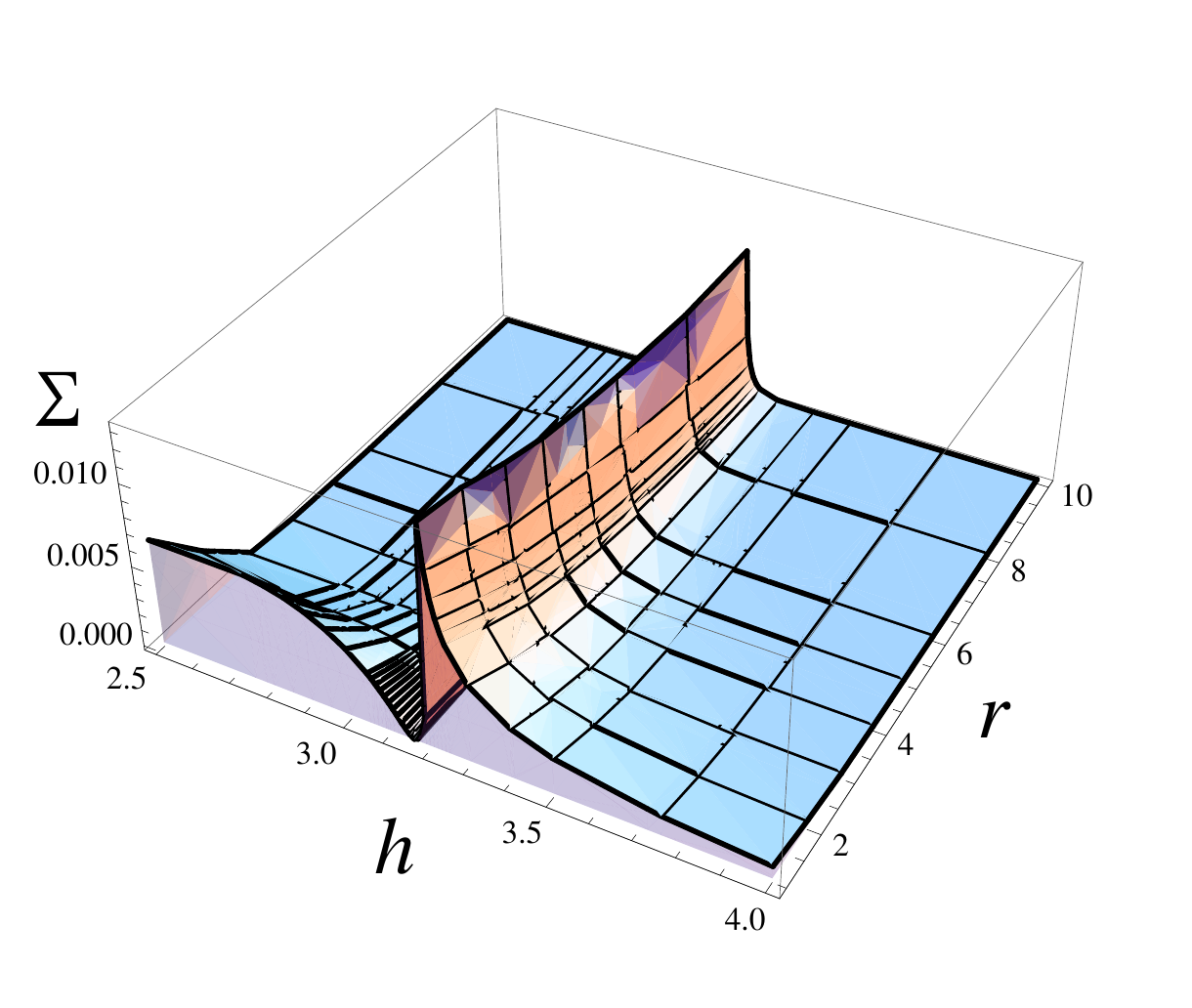}
  \caption{(color online). Quadratic difference between the optimal correlations obtained following the four different 
    measurement strategies $\hat{M}$ described above (see also Fig.~\ref{fig:Classical_Corr}): 
    $\Sigma^2 \equiv \sum_{\hat{M} } \big( {\cal C}^{(r)} - \langle {\cal C}^{(r)} \rangle \big)^2$.
    Here $\langle {\cal C}^{(r)} \rangle$ denotes the average correlation with respect to the four $\hat{M}$.
    Data are shown for the $xyx$ model as a function $h$ and for different  
    $r = \vert A-B \vert$.}
  \label{fig:scarti}
\end{figure}

A similar analysis of correlations for $r > 1$ strongly suggests that the effect of different 
measurement strategies at long ranges vanishes everywhere but close to the quantum critical points, 
where the correlation functions decays algebraically with $r$ (Fig.~\ref{fig:scarti}). 

In the disordered phase of the Ising and the $xyx$ model, 
where the order parameter $\langle \hat{\sigma}^x \rangle = 0$, as well as in the $xxz$ model,
$\theta_{opt}$ and $\phi_{opt}$ are fixed to a value independent by $h$ (see Table~\ref{table_angles}). 
By analyzing the rotated measurements $\hat{B}_k (\theta,\phi)$ it turns out 
that projective measurements can achieve a local measurement along the eigenvectors $\ket{v}$ of 
%
$\hat{\sigma}_{loc} = J_x \hat{\sigma}^x + J_y \hat{\sigma}^y + J_z \hat{\sigma}^z$ ,
%
fixed by the system Hamiltonian.
For both the $xyx$ and Ising model, optimal correlations are attained 
by using projective measurements along the $x$ axis: 
$\hat{B}_\pm (\pi/2, 0) = \frac{1}{2} (\hat{\Id} \pm \hat{\sigma}^x)$.
This reflects the $Z_2$-symmetry $\sigma^x \rightarrow -\sigma^x$ of the paramagnetic phase.
On the other hand, the operators of each optimized 4-elements POVM can be written as: 
$\hat{B}_k (\theta_{opt},\phi_{opt}) = \ket{\psi_k^{opt}}\bra{\psi_k^{opt}}$, 
where $\ket{\psi_k^{opt}}$ is of the type 
$\ket{\psi_k^{opt}} = \xi_k \sqrt{1-a_z} e^{2i\phi_{opt}} \ket{\uparrow} + \sqrt{1+a_z} \ket{\downarrow}$, 
with $\xi_k \in {\mathbb C}$. It turns out that $\ket{\psi_k^{opt}} = \ket{v}$ 
with $\phi_{opt} = 0$ and  $\xi_k = \xi^{(v)} = e^{i \arg{(a_x+ia_y)}}$  for projective measurements.
For CIC measurements $\xi_k = \xi^{(v)}$, but $\phi \neq 0$; for SIC-POVM  $\xi_k \neq \xi^{(v)}$.
We note that for large $h$, where the state is nearly fully polarized along $z$, correlations 
are vanishing and therefore measurements along any direction are optimal. 
The CIC with three-parameter optimization leads to optimal correlations in the disordered phase.

\begin{center}
  \begin{table*}
    \begin{tabular}{|c|c|c|c|}
      \hline
{}    & Ising, $h>h_c$ & $xyx$, $h>h_c$ & $xxz$, quasi-lro  \\ \hline
Proj. & $\theta_1 = \pi/2$; $\phi_1 = 0$      & $\theta_1 = \pi/2$; $\phi_1 = 0$      & $\theta = \pi/2; \forall \phi$ \\
{}    & $\theta_2 = \theta_1$; $\phi_2 = \pi$ & $\theta_2 = \theta_1$; $\phi_2 = \pi$ & {} \\ \hline
CIC  & $\theta_1 = 0$; $\forall \phi_1$     & $\theta_k = \pi$;                  & $\theta=0$; $\forall \phi$  \\
     & $\theta_2=\pi$; $\phi_2 = 0, \pi$   &  $\phi_k \approx 1.478 + k\pi/2$ & \\
     &  $\forall \theta_3$; $\phi_3 = \pi/2 + k \pi$ & & \\  \hline
SIC & $\theta_k = \pi$;  & $\theta_k = \pi$; & $\theta_{k_1} \approx 0.955$; $\phi_{k_1} = 3\pi/4 + k_1 \pi$ \\
    & $\phi_k \approx 0.393 + k\pi/2$  & $\phi_k \approx 1.152 + k\pi/2$ & $\theta_{k_2} \approx 2.185$; $\phi_{k_2} \approx 3.92 + k_2 \pi$ \\ \hline
CIC 3-par & $J_x = 1; \, J_y = J_z = 0$ & $J_x = 1; \, J_y = J_z = 0$ & $J_z = 0$; $\forall (J_x, J_y)$ \\
      \hline
    \end{tabular}
    \caption{Optimization angles $\theta_{opt} \in [0, \pi]$ and
      $\phi_{opt} \in [0, 2 \pi)$ in the disordered phases with vanishing order parameter. 
        $k$ is any integer positive number. 
        In the last line: optimization directions $J_\alpha$ for CIC measures.}
    \label{table_angles}
  \end{table*}
\end{center}

\begin{figure}[!t]
  \centering
  \includegraphics[width=0.9\columnwidth]{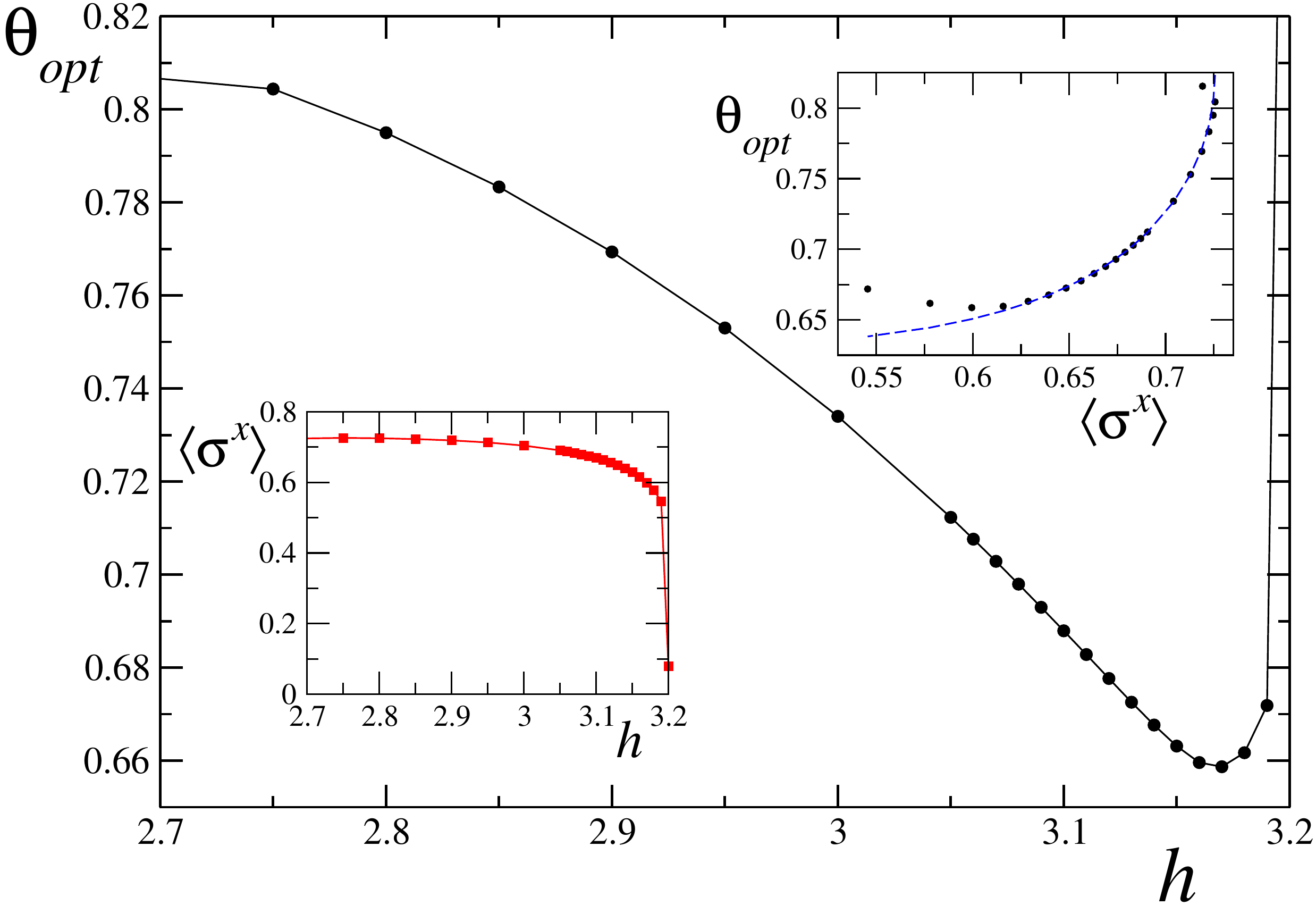}
  \caption{(color online). Optimization parameters for the nearest-neighbor correlations 
    in the $xyx$ model, after a projective measurement.
    The fit in the upper inset (dashed blue line) is $\theta_{opt} = 0.824-0.709 \sqrt{0.0769-x^8 }$.
    The other optimization strategies are discussed in the Supplementary Material.
    In the lower inset we display the order parameter $\langle \hat{\sigma}^x \rangle$
    as a function of $h$, as extracted from numerical simulations.}
   \label{bb}
\end{figure}

In the symmetry-broken phase, optimal angles $(\theta_{opt}, \, \phi_{opt})$ 
for both Ising and $xyx$ models display a non trivial dependence on the order parameter, 
as visible in Fig.~\ref{bb}. 
Except for a region close to the QPT, we fitted our results using 
the formula $\theta_{opt} = A \sqrt{B - \langle \hat{\sigma}^x \rangle^n} +k$,
where the parameters $A,B,k$ are model-dependent, while we imposed $n=8$, 
inspired by the linear variation of $\theta_{opt}$ with $h$ that we observed in the Ising model,
and by the characteristic exponent $\beta=1/8$ of its order parameter at the QPT
(see the upper inset of Fig.~\ref{bb}). 
In some cases we found significant deviations from such dependence (see Supplementary Material 
for a detailed discussion of the optimal parameters).
Peculiar behaviors arise close to the QPT, where dramatic changes appear,
and to the factorization points, where an extremal value is reached.
Such behavior is consistent with the interpretation of the factorization phenomenon 
as a ``correlation transition'' resulting from a competition between parallel 
and antiparallel entanglement~\cite{fubini2006,tomasello2011}:
Optimal correlations arise from the balance of two optimizations involving the parallel 
and antiparallel entangled components (both present in the ground state). 
The two entanglement components switch each other and an extremal optimal angle is reached at $h_f$.

It is interesting to compare the optimal angles in the $Z_2$ symmetry-broken phase with those
in the $xy$ gapless phase of the $xxz$ model, where the order parameter is vanishing in a non trivial way 
because correlations decay algebraically. For both projective and CIC measurements, the optimization 
in such phase is characterized by a fixed value of $\theta_{opt}$, $\forall \phi_{opt}$, thus indicating 
that, because of quasi-long range order, any preferential measurement direction is not unique in the phase. 
Such scenario is confirmed by the three-parameters optimized CIC POVM (last row of Table~\ref{table_angles}).

{\it Discussion.}---
We analyzed spin-spin correlations that are established after performing 
a local measurement on one of the two spins in the ground state of a quantum spin chain.
We considered projective measurements as well as symmetric IC POVMs; furthermore we engineered 
coupling-oriented IC POVMs with the aim to shed light on how the optimal measurement 
can be performed {\it a priori}, once a certain knowledge on the system has been previously acquired.
The measurement protocols were first tested regardless to adjustable parameters,
by looking at the conditional entropy.
Then we focused on the possibility to adjust the measurement on the basis of local interactions. 
Interactions, coherence and symmetry fix the ``optimal flow'' of information;
the optimal strategy to extract the correlations eventually depends on the quantum phase.

Specifically, an analysis of the quantum Ising and the $xyx$ model,
both sharing the same kind of $Z_2$-symmetry breaking QPT
(even if with very different local interactions),
showed that {\it in the ordered phase the optimal correlation follows the global order}, 
in the sense that the parameters characterizing the optimal measurement strategy vary 
with the exponent $\beta$ of the order parameter. 
Such a result could be useful to heuristically dictate 
the optimal measurement strategy also for higher order correlations (between spin-blocks), 
where the optimization protocol is not practicable.
On the contrary, {\it in the disordered phase local interactions fix the optimal strategy},
in the sense that optimal correlations are attained by fulfilling a local requirement 
of projecting along $\sigma_{loc}$. 
The results on the $xxz$ model further support this scenario:
optimal correlations are obtained for measurements respecting 
the in-plane symmetry of the model, for any fixed direction in the $xy$ plane
(there is no preferential direction because of quasi-lro in the $xy$ critical phase). 

Finally we analyzed correlations at long ranges showing that, near the QPT, 
long-range correlations are strongly affected by the measurement strategies (see Fig.~\ref{fig:scarti}). 
In the gapped phase any measurement strategy produces the same result, 
on a length-scale where the correlation functions themselves are sensible. 

Given the relation between optimal correlations and quantum discord, 
our results could be important in many-body implementations of quantum information protocols.
We also comment that, being a single-spin state fully accessible through Eq.~\eqref{eq:POVM},
our scheme provides an effective strategy to perform state tomography of one 
of the two spins (a notoriously challenging problem in quantum magnets). 

{\it Acknowledgments.}---
LA thanks the Centre for Quantum Technologies where part of this work was done.
AH is supported by the Government of Canada through NSERC and by the Province of Ontario through MRI. 
DR acknowledges financial support from EU under Grant Agreement No.~248629-SOLID.
VEK acknowledges financial support from the NSF grant DMS-0905744.


\newpage {\bf Supplementary material.}
\\

Here we report a discussion on the optimal angles $(\theta_{opt}, \, \phi_{opt})$, 
as well as on the optimal vectors $(J_x, J_y, J_z)$, which maximize the correlations 
in the symmetry-broken ordered phase. Namely, we focus on $h < h_c$ in the Ising and the $xyx$ model.
Only the case $r=1$ of two nearest-neighbor spins is addressed.

\section{Ising model}

\begin{figure}[h]
  \centering
  \includegraphics[width=0.52\columnwidth]{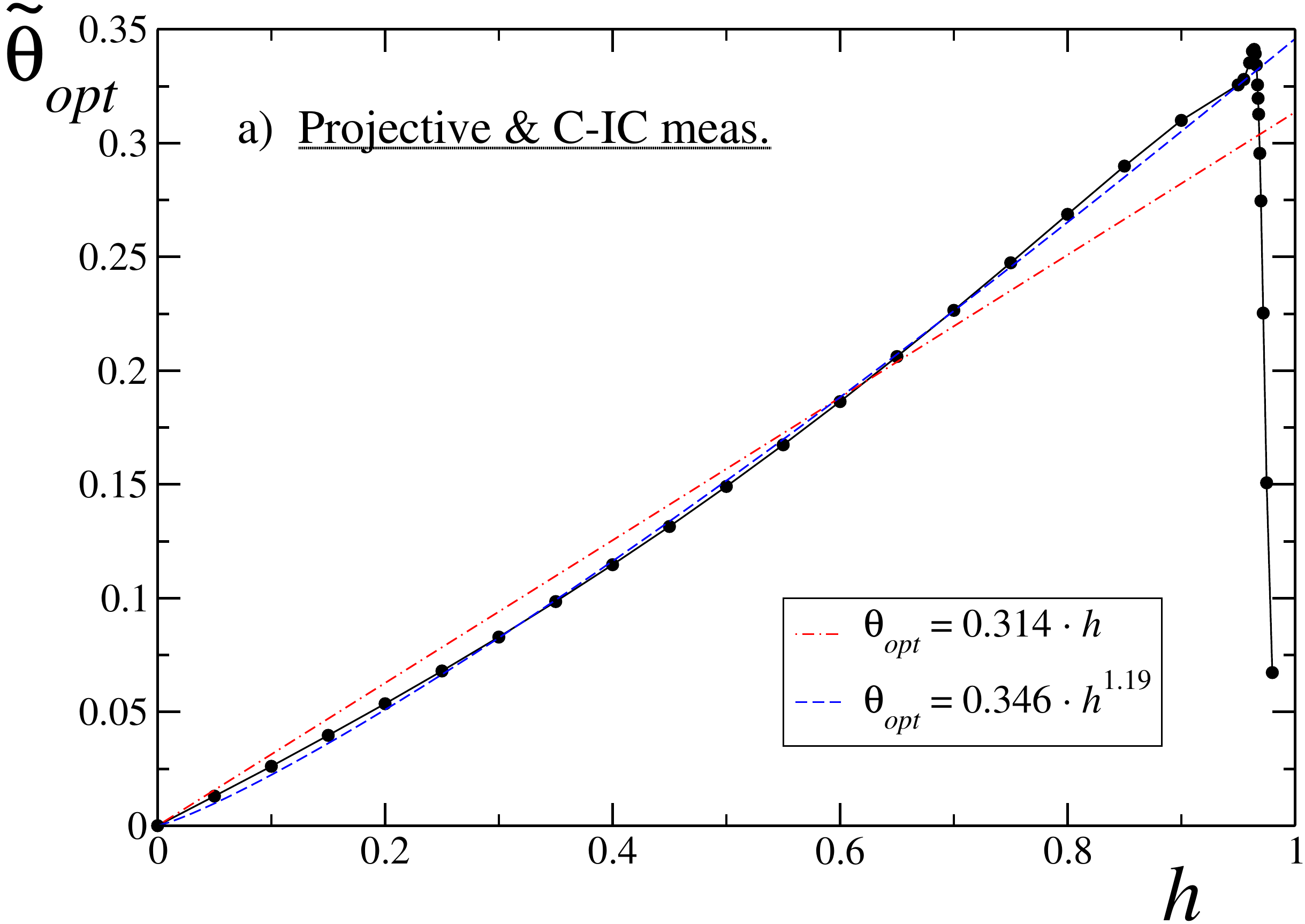}
  \bigskip
  \includegraphics[width=0.52\columnwidth]{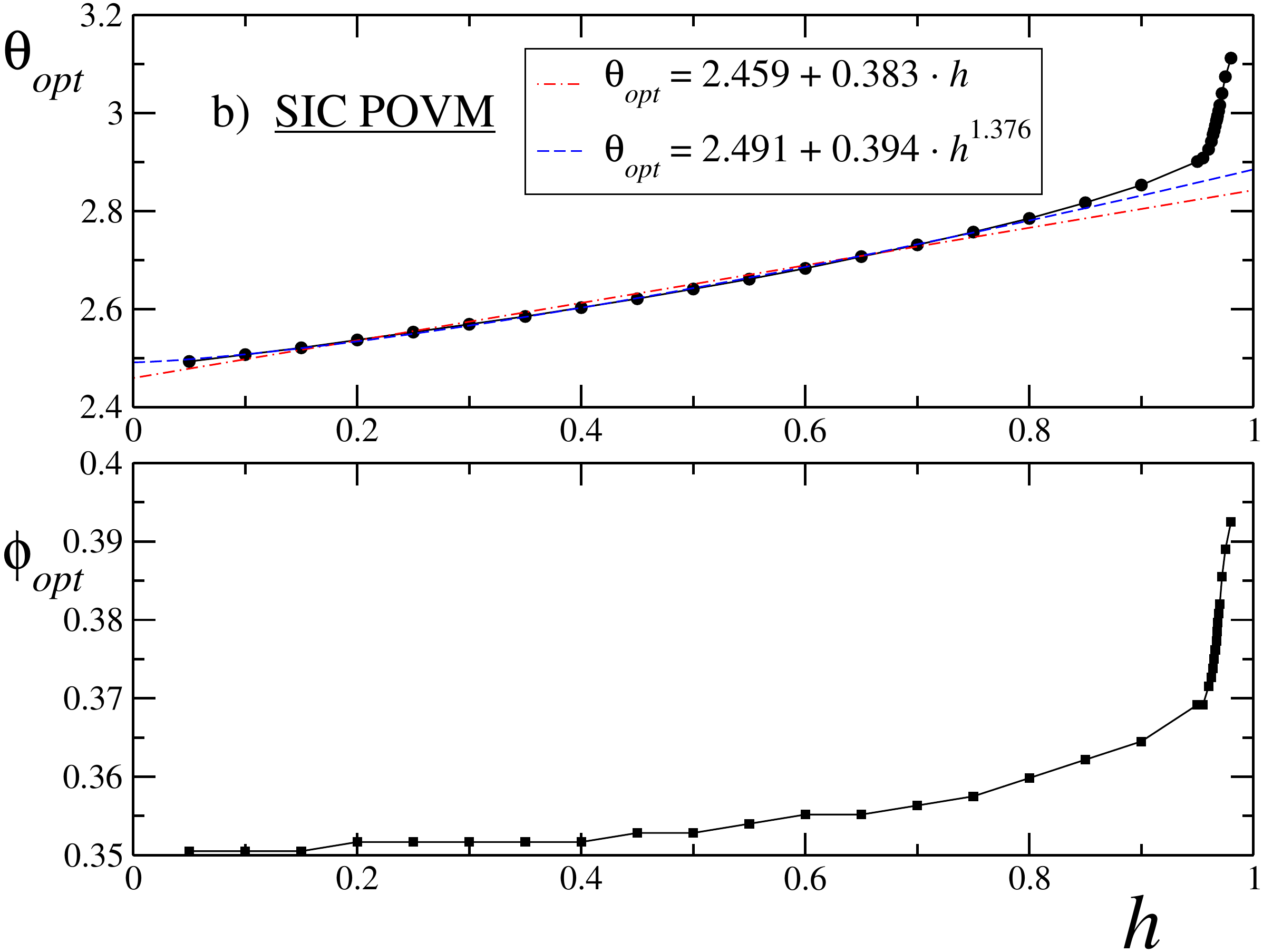}
  \bigskip
  \includegraphics[width=0.52\columnwidth]{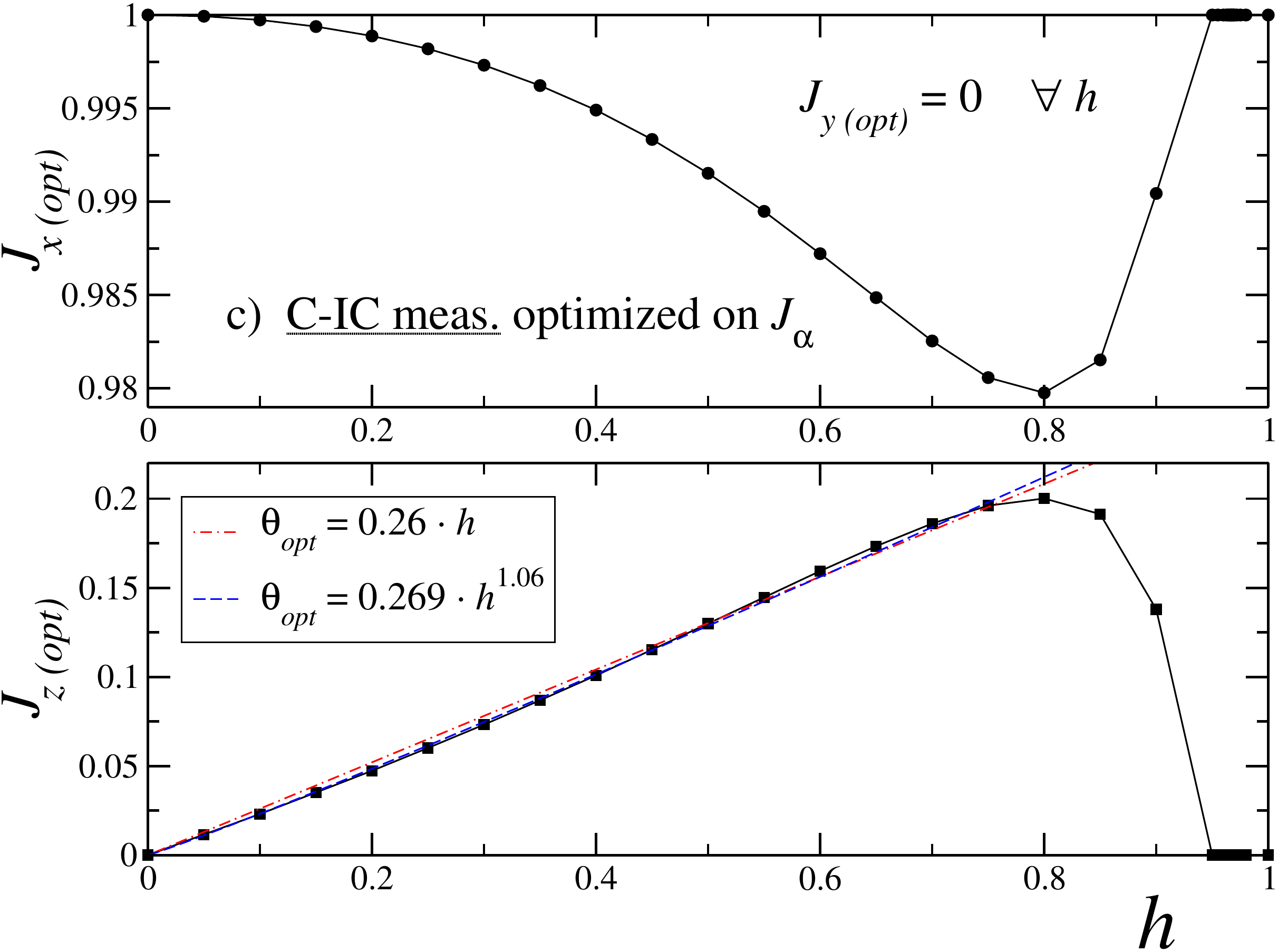}
   \caption{Optimization parameters for the nearest neighbor correlations in the Ising model.
     The different panels correspond to different optimization strategies, as discussed
     in the main text. Red dotted-dashed lines are linear fits of data, while blue
     dashed ones are power-law fits; fitting parameters are indicated in the respective legends.
     The tilde in panel a) stands for a rescaling of $\theta_{opt}$ such that $\tilde{\theta}_{opt}(h=0) = 0$.}
   \label{fig:Ising_opt}
\end{figure}

For both projective and C-IC measurements there are two optimal couples of angles. 
The first pair is determined by $\phi_{opt}^{(1)} = 0$ and $\theta_{opt}^{(1)}(h=0) = \pi/2$,
with $\theta_{opt}^{(1)} \propto h$ (Fig.~\ref{fig:Ising_opt}a), as long as $h$ increases until 
the critical point $h_c = 1$ (note however the dramatic changes in $\theta_{opt}$ close to $h_c$).
The other pair of optimal angles is $( \pi - \theta_{opt}^{(1)}; \pi )$. 
The approximately linear behavior of $\theta_{opt}$ with respect to the field $h$
results in a fitting law analogous to Eq.~(6) in the main text, namely:
\begin{equation} 
  \theta_{opt} =A \sqrt{B - \langle \hat{\sigma}^x \rangle^n} +k \; .
  \label{eq:theta_opt_ising1}
\end{equation} 
In the specific we found $A \approx 0.34$ and $B=1$; this is because, for the Ising model, 
$\langle \hat{\sigma}^x \rangle = \vert 1 - h^2 \vert^{1/8}$.

For SIC POVM  $\theta_{opt}$ displays corrections to the linear behavior
with $h$, while $\phi_{opt} \approx const.$, as compared to the scale of
variation of $\theta_{opt}$ (Fig.~\ref{fig:Ising_opt}b).  
We point out that there are other optimal points which maximize the correlations
to the same amount of the ones depicted in the figure (up to numerical precision).
Here we select this point for continuity with the $(\theta_1 = \pi; \, \phi_1 \approx 0.393)$ 
optimal point in the disordered phase [see Table~(1) in the main text].

For C-IC optimized on the three directions $(J_x, J_y, J_z)$, 
the optimal point is found to be at $J_y = 0$, while $J_x$ and $J_z$ vary 
with $h$ according to Fig.~\ref{fig:Ising_opt}c.
Away from the critical point, we can fit $J_z$ linearly with $h$.
Note that the value of $J_x$ is fixed by the normalization constraint $\vert\vert \vec{a}_k \vert\vert = 1$,
and its variation is much smaller on the scale of $J_z$.

\section{xyx model}

As in the Ising model, for the projective measures we found two optimizing points.
The dependence of $\theta_{opt}$ with $h$ for one of the two points is depicted in Fig.~(5) of the main text,
while the corresponding axial angle $\phi_{opt}^{(1)}=0$, independently of $h$. 
The other pair of optimal angles is $( \pi - \theta_{opt}^{(1)}; \pi )$, 
in complete analogy with the Ising model.
We fit $\theta_{opt}$ as a function of the order parameter $\langle \sigma^x \rangle$ 
as in Eq.~\eqref{eq:theta_opt_ising1}; the fitting parameters are $A = −0.709$ 
and $B = 0.0769$ [see dashed blue curve in Fig.~(5) of the main text].

\begin{figure}[h]
  \centering
  \includegraphics[width=0.52\columnwidth]{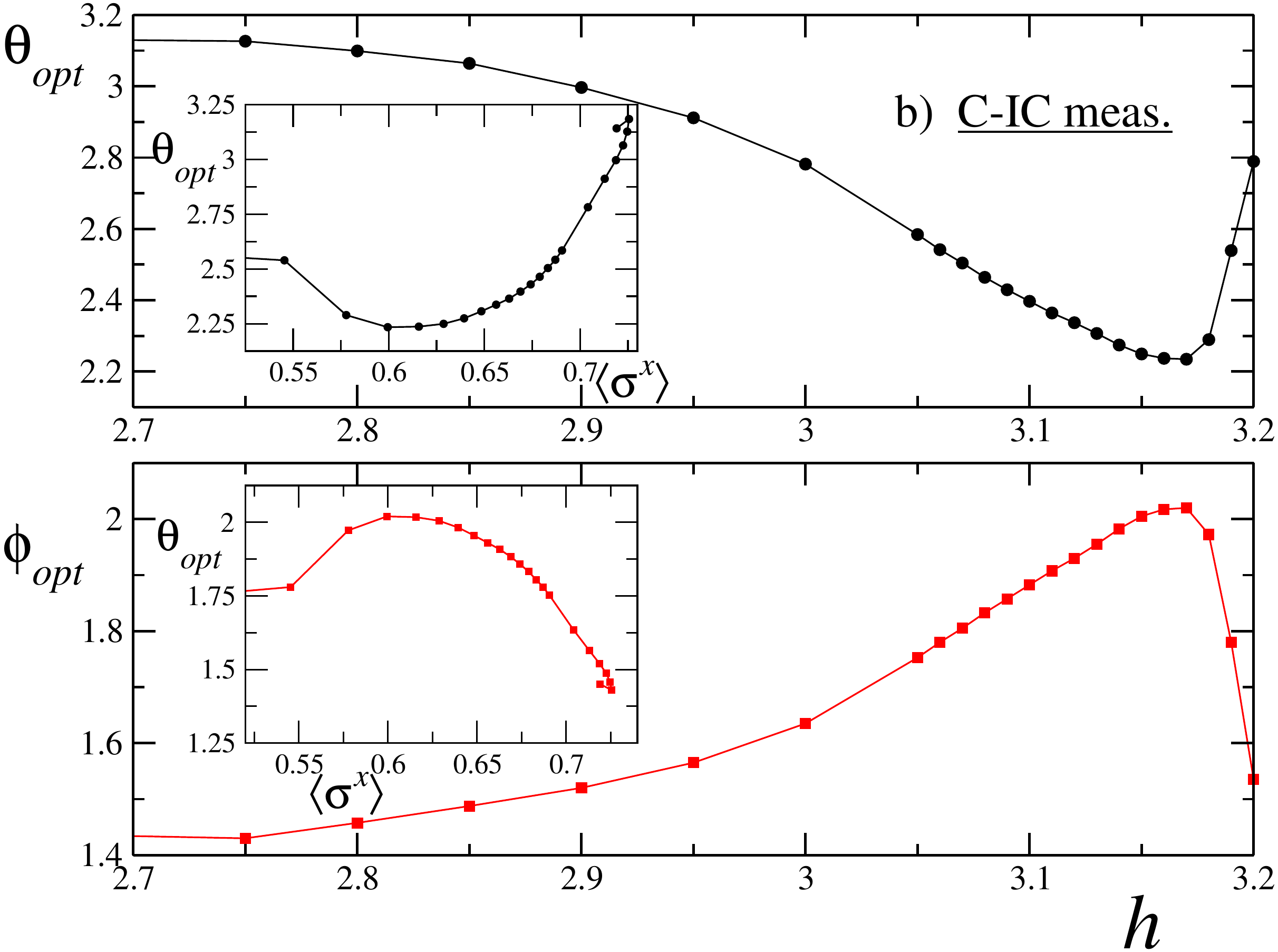}
  \bigskip
  \includegraphics[width=0.52\columnwidth]{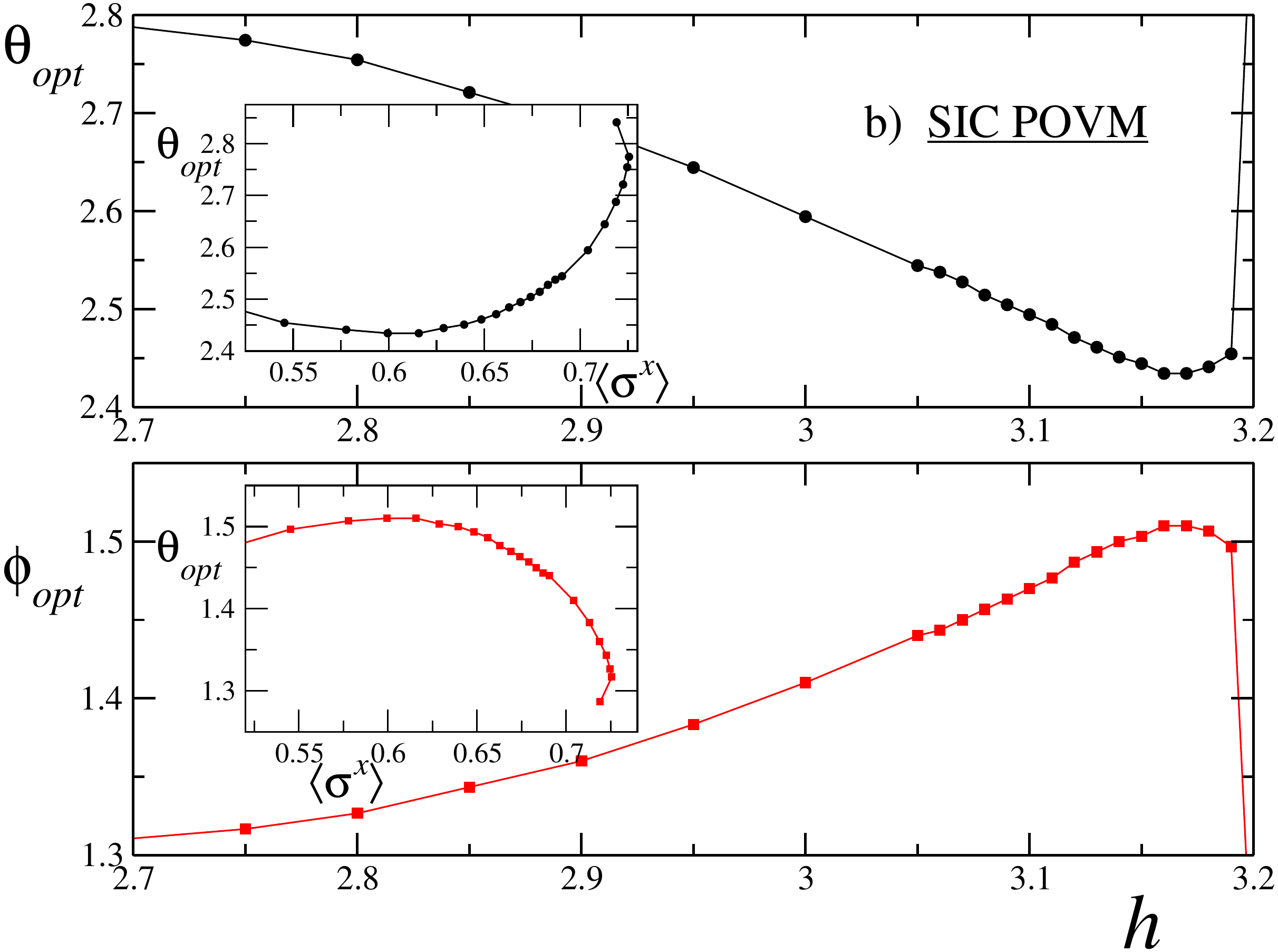}
  \bigskip
  \includegraphics[width=0.52\columnwidth]{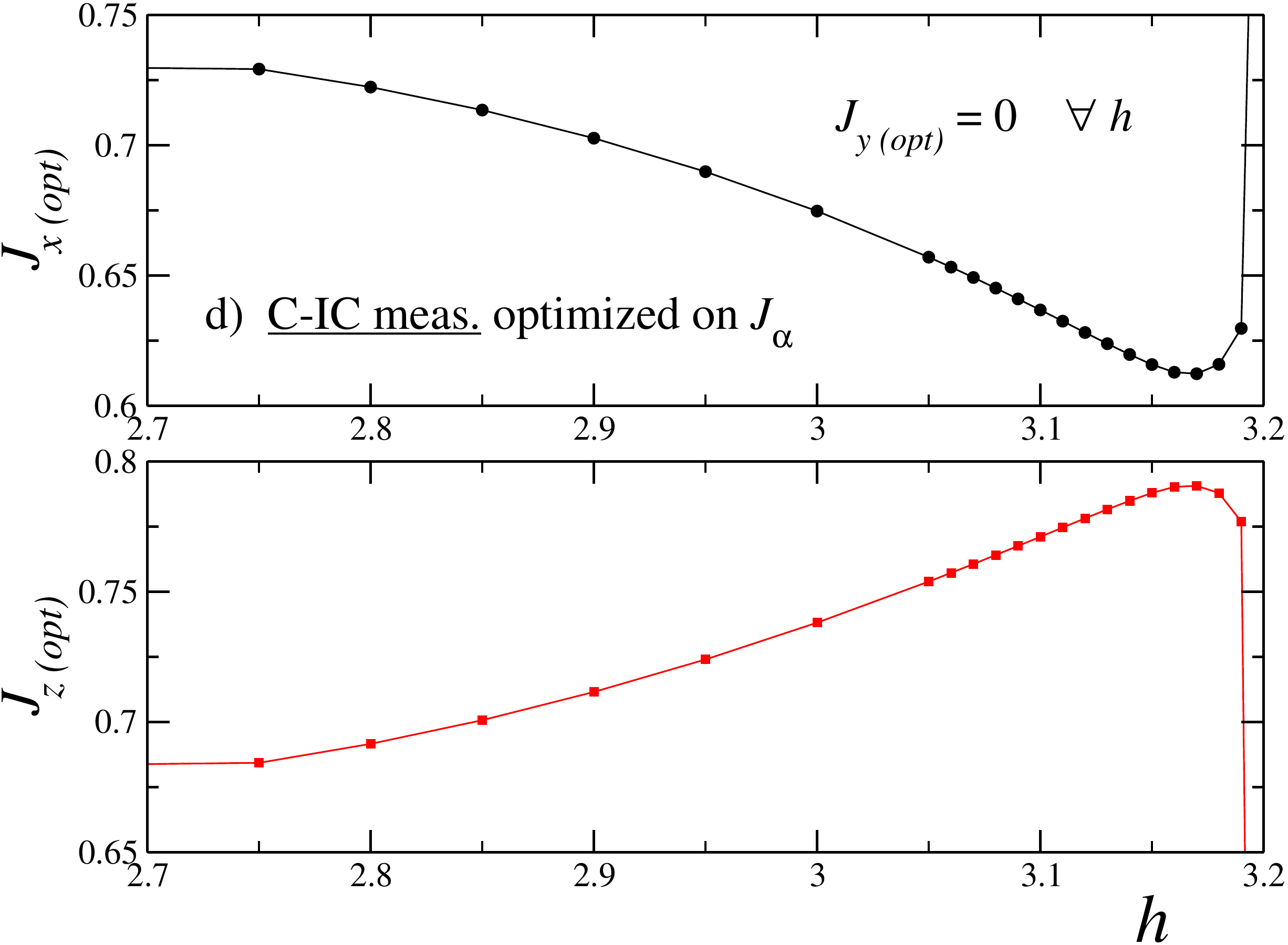}
   \caption{Optimization parameters for the nearest neighbor correlations in the $xyx$ model.
     The different panels correspond to different optimization strategies, as discussed in the main text.
     The insets display how the optimal angles are pronounced non-linear functions of $\langle \sigma^x \rangle$.} 
   \label{fig:xyx_opt}
\end{figure}

For the other measurement strategies we observed deviations from this behavior
and we could not operate such a fit. 
The dependence of the optimal angles on $h$ is displayed in the various panels of Fig.~\ref{fig:xyx_opt}.

\widetext

\begin{thebibliography}{99}
\bibitem{note}
  For spin-spin correlations the optimization in Eq.~\eqref{eq:Class} 
  can be constrained to 4-projectors-elements POVMs~\cite{dariano}. 
  Though von Neumann measurements were proved to be most effective than 3-element POVMs, 
  there is no conclusive proof for the efficiency of 4-elements POVMs~\cite{zaraket}. 
  The optimization within 4-elements POVMs would involve eleven raw parameters.
\end{thebibliography}
\end{document}